\documentclass[preprint2]{aastex}

\newcommand{\phiprec}{\mbox{$\phi_{\mbox{\tiny 35}}$}}
\newcommand{\phiorb}{\mbox{$\phi_{\mbox{\tiny orb}}$}}

\newcommand{\her}{\mbox{Her~X-1}}
\newcommand{\chired}{{\mbox{$\chi^2_{\mbox{\tiny red}}$}}}

\newcommand{\xte}{\mbox{\em RXTE}}

\newcommand{\etal}{\mbox{et\ al.\ }}
\newcommand{\dex}[1]{\hbox{$\times\hbox{10}^{#1}$}}

\newcommand{\kmsec}{\,\mbox{$\mbox{km}\,\mbox{s}^{-1}$}}

\newcommand{\kev}{\,\mbox{keV}}

\newcommand{\msun}{\,\mbox{$\mbox{M}_{\odot}$}}

\slugcomment{Submitted to the Astrophysical Journal}

\shorttitle{Reflection from Hercules X-1}
\shortauthors{M.\,D.\ Still et~al.}

\begin{document}

\title{Atmospheric reflection during an anomalous low-state of 
Hercules~X-1}

\author{Martin Still\altaffilmark{1}}
\affil{NASA/Goddard Space Flight Center, Code 662, Greenbelt, MD~20771\\
Physics and Astronomy, University of St Andrews, North Haugh, 
St Andrews, Fife KY16~9SS, UK}

\author{Kieran O'Brien, Keith Horne}
\affil{Physics and Astronomy, University of St Andrews, North Haugh,
St Andrews, Fife KY16~9SS, UK}

\author{Bram Boroson, Lev G. Titarchuk, Kimberly Engle\altaffilmark{2}}
\affil{NASA/Goddard Space Flight Center, Greenbelt, MD~20771}

\author{Saeqa D. Vrtilek}
\affil{Harvard-Smithsonian Center for Astrophysics, 60 Garden Street, 
Cambridge, MA 02138}

\author{Hannah Quaintrell}
\affil{Department of Physics, The Open University, Milton Keynes 
MK7 6AA, UK}

\author{Hauke Fiedler}
\affil{Institut of Astronomy and Astrophysics, Ludwig-Maximilian 
University, D-81679 Munich, Germany}

\altaffiltext{1}{Universities Space Research Association}
\altaffiltext{2}{SSAI, 5900 Princess Garden Parkway, Suite 300, Lanham
MD 20706}

\begin{abstract} 
We  present \xte\ observations of the  eclipsing X-ray binary Hercules
X-1 conducted during an anomalous low state.  Data reduction reveals a
light curve over 2.7 orbital  cycles remarkably similar to optical and
UV light  curves  which are dominated by   the companion star.   Count
rates are modulated close  to the orbital  period, attaining a maximum
when the inner  face of the companion  star, irradiated by X-rays from
the compact  source, is  most-visible.   Cold reflection   provides an
acceptable fit  to the energy  spectrum.  Employing binary geometry to
scale the model and assuming companion star reflection, we are able to
reconstruct the   incident X-rays which are   removed  from our direct
line-of-sight (presumably by the accretion disk).  We find the flux of
the hidden source to be identical to the observed flux of \her\ at the
peak  of its  main-high  state.  Consequently,   \her\  is  emitting a
reflected spectrum,   largely uncontaminated by  direct X-rays  in the
anomalous  low-state.    The  spectral   energy  distribution, period,
amplitude  and  phasing of the  modulation  are all consistent  with a
companion star origin.  Since this source  occurs in a well-understood
binary  environment,   it provides  an excellent   case study for more
sensitive experiments in the future.
\end{abstract}


\section{Introduction}

Her~X-1 (Tananbaum \etal 1972) is an eclipsing X-ray binary containing
a pulsar of 1.4  \msun\  and an  A7  stellar companion of  2.2  \msun\
(Middleditch \& Nelson 1976; Reynolds \etal 1997). The system displays
behaviour  on four separate  periods -- the  pulsar spin (1.24-s), the
binary orbit (1.7-d),  a super  period of  35-d  which results from  a
retrograde-precessing,  warped accretion disk,  and a beat between the
precessional and orbital  periods of 1.62-d.   The warp phenomenon  is
poorly understood but its source is likely to be a combination of both
tide- and radiation-driven  precession  (Papaloizou \&  Tarquem  1995;
Pringle 1996).

The 35-d  super-cycle results in two  phases of  strong X-ray activity
per  cycle (Giacconi \etal 1973;  Scott \& Leahy 1999).  The main-high
state has a rapid turn-on  over $\sim$90-m and decays over $\sim$10-d.
An  off-state where flux  remains at $\sim$1  percent of the main-high
state  level  follows for the next   $\sim$10-d, succeeded by a second
short-high state, lasting  $\sim$5-d,  with flux peaking  at  $\sim$30
percent  of maximum.  A    further off-state completes the  cycle  and
extends over the  next $\sim$10-d.  The accretion  disk casts a shadow
over the surface of the companion  star which migrates across the star
on the beat frequency.  The super-cycle is consequently observed at UV
and optical energies which are dominated by X-ray emission reprocessed
in the stellar atmosphere (Gerend \& Boynton 1976).

The  35-d     clock   has  remained    mostly   coherent   since  it's
discovery. There have however been three  occasions when the clock has
missed    several consecutive turn-ons,   or  main-high  flux has been
reduced  (Parmar \etal 1985; Vrtilek   \etal 1994; Parmar \etal 1999).
The  cause of these anomalous   low-states is not clear, but  probably
result from changes in the state of the accretion disk, either through
an   increase in  vertical  scale  height,  or  disruption to the disk
warp. However  the level of  reprocessed UV and  optical flux from the
companion star during these  states suggests that the intrinsic  X-ray
flux, and presumably the accretion  rate, remains mostly unaffected by
this event  (Vrtilek \etal 1994; 2001).   The high orbital inclination
of the binary  indicates  that perhaps just  a small  variation in the
disk geometry  will block direct X-rays  from our line-of-sight, while
the  companion stars view remains  largely unchanged over  much of its
surface area.

The latest anomalous  low-state  began during 1999  Feb  (Parmar \etal
1999) and the  on-states had not  recovered four cycles later when the
current observations   were   scheduled.    These were   part  of    a
simultaneous multi-wavelength  campaign on \her\  described by Vrtilek
\etal (2001).  Results from simultaneous Hubble Space Telescope orbits
are detailed by Boroson \etal (2000) and  Boroson \etal (2001), and an
earlier epoch of short-high  state visits with  {\xte} by  Still \etal
(2001).

The  current  \xte\ data reveal  many similarities  between  the X-ray
light curve and  its optical/UV counterparts,  suggesting that much of
the anomalous low  state X-ray  emission could  be of  companion  star
origin.   We find that  cold   Compton reflection  is  a statistically
acceptable model for the energy spectrum and  a reflection surface the
size and distance from the  pulsar of the  companion star gives a good
prediction of  the incident X-ray flux provided  this is  identical to
the peak of the normal on-state intensity.   The possible detection of
a relatively  clean reflection  spectrum  in a well-determined  binary
geometry, free of dominant direct X-rays, would be a valuable resource
for    investigators   of   reflecting     atmospheres  in   the  less
well-determined environments of AGN  and stellar black hole candidates
(Lightman \& White; White, Lightman and Zdziarski 1988).

\section{Observations}

\begin{figure*}
\begin{picture}(100,0)(10,20) 
\put(0,0){\includegraphics{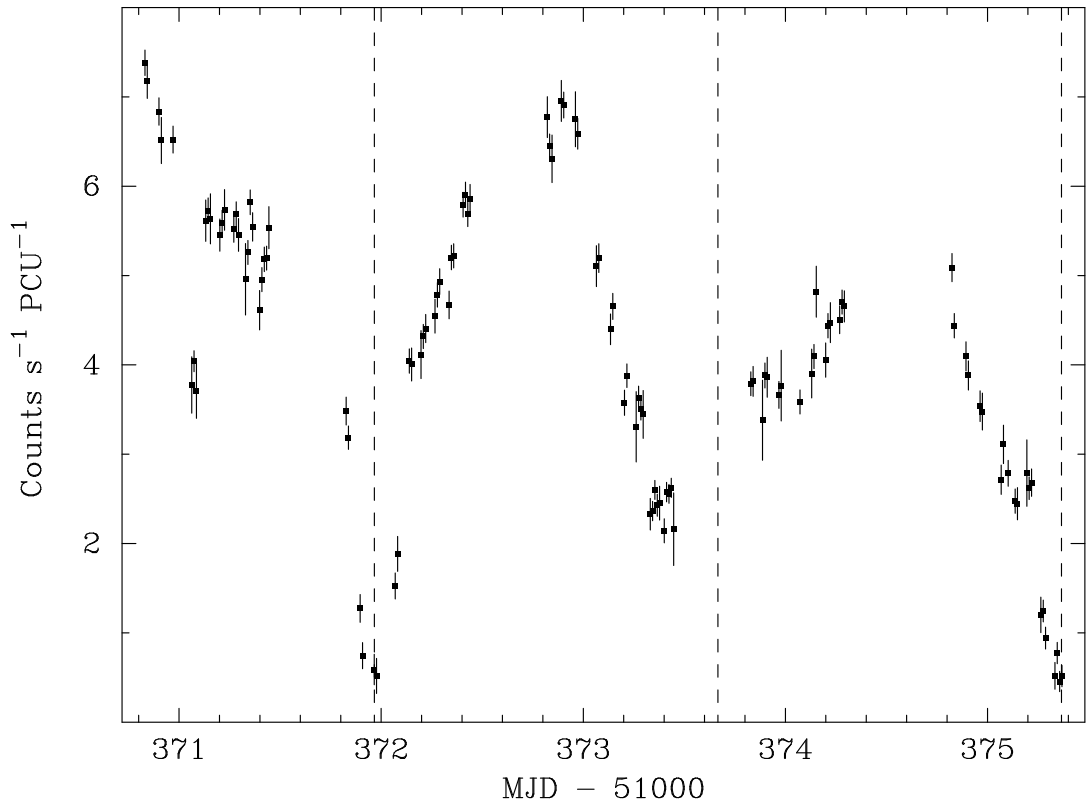}}
\noindent
\end{picture}
\vspace{90mm}
\figcaption[f1.ps]{  The 3--30 keV band light curve over the duration
of the observations. The vertical  dashed lines correspond to superior
conjunction of the neutron star, according to the orbital ephemeris of
Deeter \etal (1991). \label{lc}}
\end{figure*}

\xte\  pointed  at  \her\  intermittently  between   1999 July  11--16
(MJD~51\,370.83--51\,375.37)  accumulating  100-ksec   of  time-tagged
events.  Event reconstruction  was performed using standard algorithms
within  {\sc  ftools} v4.2.   We  analyze  data from  the Proportional
Counter Array (PCA;  Zhang \etal  1993)   in which three  of the  five
identical Xe Proportional Counting Units  (PCUs) were operational with
a combined effective  area of  3900 cm$^2$.  As  well  as the standard
data  formats, we  employ two event  analyzers  in the GoodXenon event
mode with 2-s  readout -- timestamps are  resolved  to 1-$\mu$s, while
employing the full 256   energy channels (2.2--119.0~keV during  epoch
4).  Response matrices from 1999 Mar 30 were obtained from the HEASARC
archive\footnote{http://xte.gsfc.nasa.gov}.  Background estimates  are
derived            from       the       Very        Large        Event
model\footnote{http://lheawww.gsfc.nasa.gov/$\sim$stark/pca/pcabackest.html}
to  account for cosmic events, internal  particle generation and South
Atlantic Anomaly activation. Events  from the High Energy X-ray Timing
Experiment (HEXTE), also onboard  the spacecraft, proved too scarce to
be useful.

Throughout this paper we will  employ the orbital ephemeris of  Deeter
\etal (1991):
\[
T_{\mbox{\tiny orb}} = 
MJD\,43\,804.51998(1) + 1.70016772(1) E_{\mbox{\tiny orb}}
\]
\begin{equation}
- 5.2(5) \dex{-11} E_{\mbox{\tiny orb}}^2
\label{eqn:orbit}
\end{equation}
where $T_{\mbox{\tiny  orb}}$ corresponds  to superior  conjunction of
the compact  star and   $E_{\mbox{\tiny orb}}$  is the  orbital  cycle
number. Assuming the 35-d epoch is determined by:
\begin{equation}
T_{\tiny 35} = \mbox{MJD}\,50\,041.0 + 34.85\,E_{\tiny 35} 
\end{equation}
where $T_{\tiny 35}$ corresponds to X-ray turn-on of the main-high state
and  $E_{\tiny   35}$   is  the   cycle  number   (M.  Kunz,   private
communication),  the  precession  phases  sampled    are  $\phiprec$ =
0.16--0.29.  These would  normally correspond approximately to the the
second half of a main-high state (Scott and Leahy 1999), but the average
count rate of 4 counts s$^{-1}$ PCU$^{-1}$  testifies to the anomalous
extended low-state in which \her\  had remained for the previous  four
35-d cycles (Coburn \etal 2000). Count rates are of order 1 percent of
the normal main-high state (dal Fiume \etal 1998).

\section{Light curve}
\label{sec:lc}

\begin{figure*}
\begin{picture}(100,0)(10,20) 
\put(0,0){\includegraphics{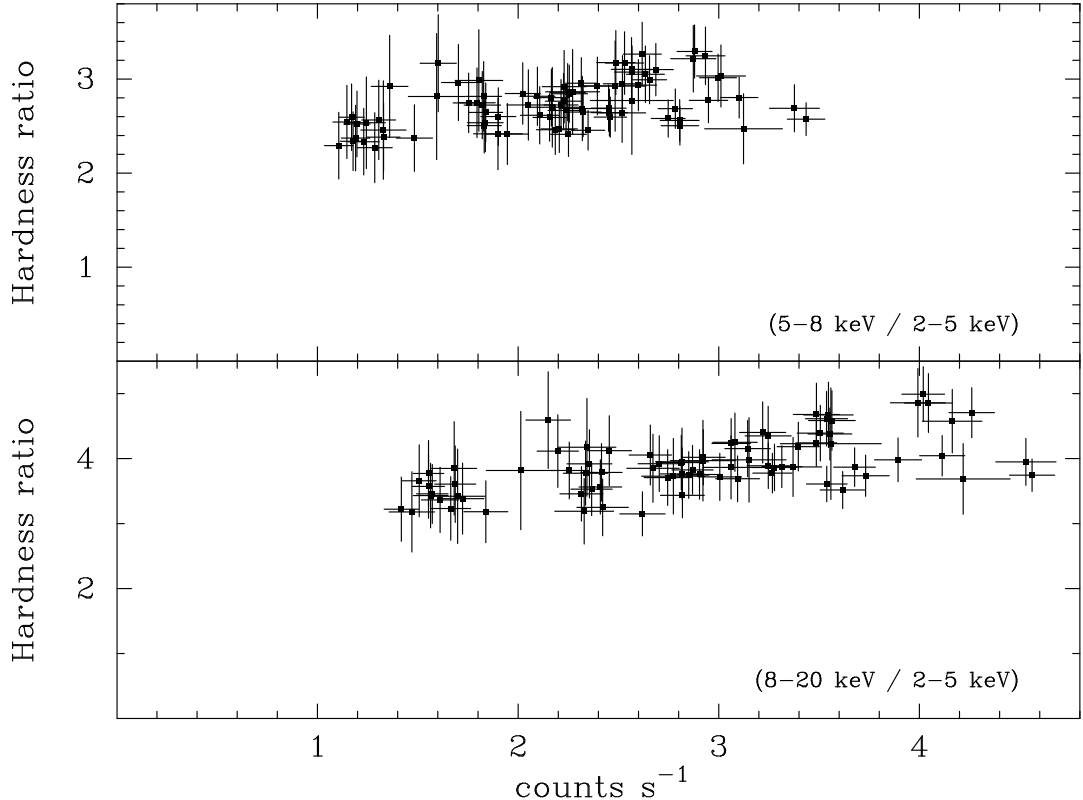}}
\noindent
\end{picture}
\vspace{95mm}
\figcaption[fig2.ps]{  Hardness ratios for the bands 2--5 keV, 5--8 keV 
and   8--20 keV.  Eclipse points,  \phiorb\   = $-$0.1--0.1, have been
excluded. \label{hardness}}
\end{figure*}

Standard2  data  from each  visit   were filtered to reject  pointings
closer than 10 degrees to the Earth's  limb and those off-axis by 0.02
degrees or greater.   From all  PCA layers  and columns,  events  were
summed across  pulse-height channels to provide  a light  curve in the
energy  range 3--30\kev\ with 1-ksec sampling.    This is presented in
Fig~\ref{lc}  and  hardness   ratios between  2--5\kev,  5--8\kev\ and
8--20\kev\ are provided  in Fig~\ref{hardness} (eclipse data have been
removed from  the  hardness  diagrams).  Background  models  have been
subtracted and data divided by the number of active PCUs.

Gerend \&   Boynton  (1976) show  that  the   majority of power   from
companion star emission occurs on  the 1.7-d orbital frequency through
stellar rotation.  However, light  curve structure associated with the
accretion  flow from the  companion star,  such  as  the hot spot  and
splash from the collision of stream and  disk (Schandl 1996), occur on
a timescale close to the beat  period. Fig~\ref{lc} shows an intensity
modulation approximating  the 1.7-d orbital  period.  Depending on its
source,  we  expect the observed  modulation   to occur either  on the
orbital or 1.62-d beat period.

\begin{figure*}
\begin{picture}(100,0)(10,20) 
\put(0,0){\includegraphics{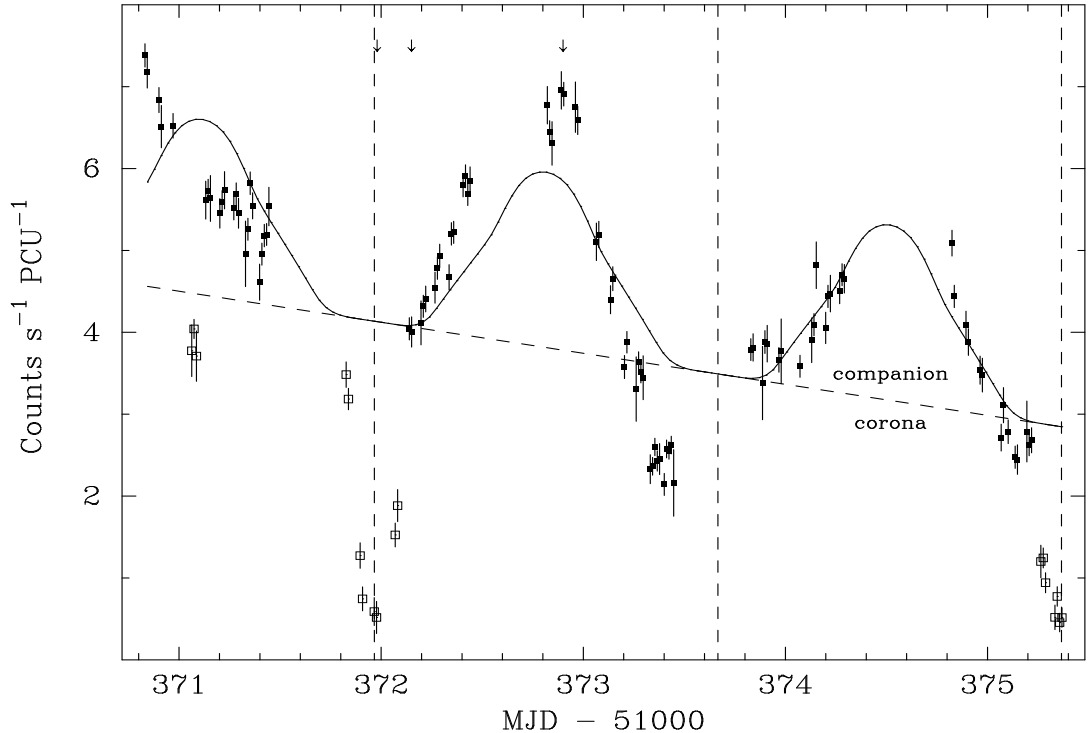}}
\noindent
\end{picture}
\vspace{85mm}
\figcaption[fig3.ps]{A fit to the light curve using the geometric
binary  synthesis code of    Still  \etal (1997) plus a   linear-decay
function to represent the corona.  Points  designated by open squares,
corresponding to eclipses and the  dip at MJD\,51,371.05 were  ignored
by the  fit.   The dashed  line is the  coronal fit   alone. The three
spectra presented in Fig.~4  were extracted from the visits  indicated
by arrows.\label{lcfit}}
\end{figure*}

The  hardness ratios   of    Fig.~\ref{hardness}  indicate  that   the
modulation  is not the   result of photo-electric absorption.   We are
observing   intrinsic flux increases or  opaque  obstruction of a soft
emission  region.   The   observed  structure  occurs  over timescales
several factors  longer than normal   main-high dipping events  (Leahy
1997).

Maximum   flux  corresponds    approximately    with   orbital   phase
$\phi_{\mbox{\tiny orb}} = 0.5$, where $\phi_{\mbox{\tiny orb}} = 0.0$
is superior conjunction of the neutron star.   This is consistent with
a region associated with the irradiated inner surface of the companion
star.  The similarities between  this light curve  and the optical, UV
and  EUV light curves which  are dominated by  companion star emission
are striking (Still \etal 1997; Leahy \&  Marshall 1999; Boroson \etal
2000).    We suggest  that the  modulation  is  the result of  Compton
reflection of  X-rays off the atmosphere of  the companion,  which was
predicted by  Basko, Sunyaev \&  Titarchuk (1974) and detected  at EUV
wavelengths by Leahy \& Marshall (1999) and Leahy \etal (2000).

If we are detecting companion star reflection, there  must be a second
source of X-ray flux  associated with the compact  object in order  to
produce the   narrow  eclipses at MJD\,51\,372.0,   MJD\,51\,373.7 and
MJD\,51\,375.4.    There are equivalent  eclipses   at optical and  UV
wavelengths which are thought to be  the eclipse of  disk light by the
companion star (Gerend \& Boynton 1976; Boroson \etal 2000).

After removing eclipses  and using the Lomb-Scargle statistic (Scargle
1982), the best fit period  to this light  curve is $1.74 \pm 0.06$-d,
with a false   alarm-probability   of $10^{-12}$.  The error    is the
$\sigma$-width of the   peak in  the   power  spectrum  and can   only
discriminate between the orbital and beat frequencies to 1.3-$\sigma$.

We  attempt a  crude  fit  to  the  light curve where   the reflection
component is constructed  by combining the geometric stellar synthesis
code of Still \etal (1997)  with  the classical H-function  reflection
coefficients  (Chandrasekhar 1960).  Reflected flux  between  3 and 30
\kev\ is integrated over the stellar surface area.  This is detailed in
Appendix~A.  We adopt an orbital   inclination of $i$ = 82$^\circ$,  a
neutron  star mass of $M_{\mbox{\tiny  ns}}$  = 1.4\msun\  and a Roche
lobe-filling companion star mass of $M_{\mbox{\tiny com}}$ = 2.2\msun\
(Reynolds  \etal 1997).   Unlike Still  \etal  (1997), and in order to
reduce the number of free  parameters, we do not  attempt to model the
shadow of  the accretion disk over the  companion star, or the orbital
eclipse of  the  star by the disk,  centered  at \phiorb\  = 0.5.  The
energy  spectrum of the compact  source is taken to  be  a powerlaw of
slope $\alpha  =  0.9$ with  an exponential  cutoff at $E_{\mbox{\tiny
cut}} =  11$ \kev\ (see  Sec.~\ref{sec:spectrum}).  The  luminosity of
the  source is arbitrary,  where  the resulting reflection spectrum is
rescaled such that the maximum count rate at \phiorb\ = 0.5 is 1 count
s$^{-1}$ PCU$^{-1}$.  The coronal component is represented by a linear
function which is better statistically than  a constant, such that the
overall fit is given by:
\begin{equation}
L(t) = p_1 + p_2 t + p_3 R(t) 
\label{eq:lc}
\end{equation}
$p_1$ and $p_2$  are  the constant   and linear terms  of the  coronal
component,  and $p_3$ is   a renormalization constant for the  modeled
companion  star light  curve $R(t)$. $t$   is  time in  days after MJD
51,000.

After rejecting eclipse points and the dip at MJD\,51\,371.05, the fit
was made using a  Downhill Simplex Algorithm  (Press \etal 1992).  The
result is  plotted  in Fig.~\ref{lcfit}  where $p_1$   = 145, $p_2$  =
$-$0.38 and  $p_3$  = 2.14; $\chi^2$  = 2100  for  84 dof.  Decreasing
coronal  flux is consistent with the  behaviour of direct photons from
the accretion region during this phase of the normal 35-d cycle (Scott
\& Leahy 1999).

The fit is poor, although this  is not particularly discouraging given
our first-order    treatment of coronal   variability.  Residuals  are
probably dominated   by the   coronal  component.  In   the  following
sections we perform spectral fits  in an attempt  to separate the  two
components, test the suitability of reflection  for fitting the energy
spectrum of   the  modulated component, and  subtract  modeled stellar
reflection spectra to reveal the spectral variability of the corona.

\section{Energy spectrum}
\label{sec:spectrum}

\begin{figure*}
\begin{picture}(100,0)(10,20) 
\put(0,0){\includegraphics{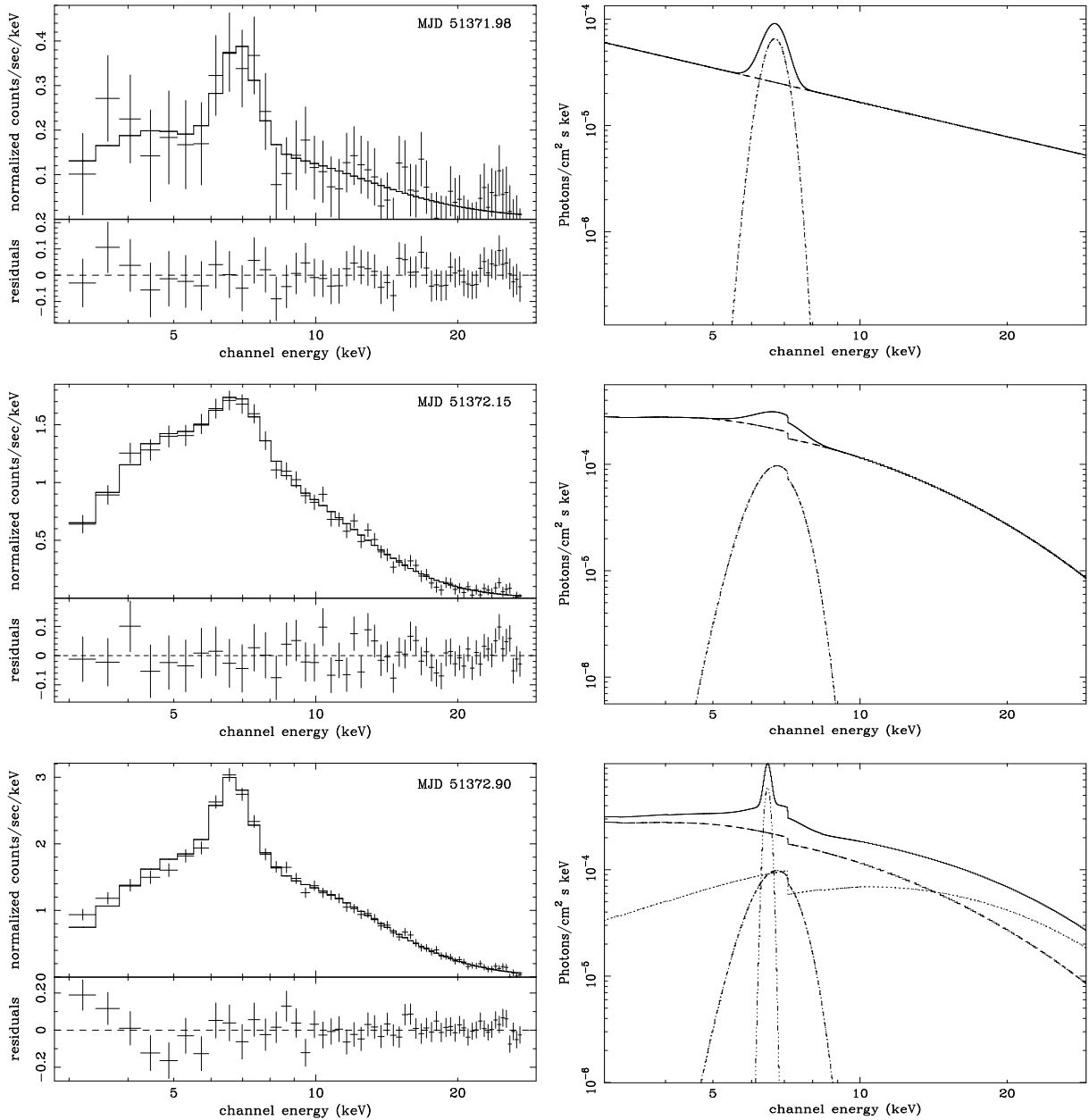}}
\noindent
\end{picture}
\vspace{160mm}
\figcaption[fig4.ps]{ Spectral fits from three visits to \her. The top 
left panel is a  powerlaw + Gaussian profile  fit to  mid-eclipse data
(MJD 51371.98). The right hand panel presents  the best fit model. The
middle  two panels are the corresponding  fit and  model of the corona
for  the  post-eclipse    visit  at MJD~51372.15  consisting    of  an
exponentially-cutoff   powerlaw  and   line,  absorbed  by  a  partial
cover. The  lower panels  represent  the visit  at  maximum  flux, MJD
51372.90, where a reflection component and second line are added.  See
Sec.~4 for details of the spectral models. \label{spectra}}
\end{figure*}

Under the same filtering constraints   as above, the GoodXenon  events
were   sampled   with the full  pulse-height   resolution  of the PCA.
Channels 1--7 ($<$  3 keV) were  ignored  due to uncertain  background
modeling, and channels  63--256 ($>$ 30   keV) ignored as a result  of
poor counting statistics.  {\sc xspec}    v11 was used at all   times.
Spectral   fits  to    three    of   the  visits   are   presented  in
Fig~\ref{spectra}.

\subsection{Corona spectrum}

We consider a model where emission originates from two distinct binary
regions, an  accretion disk corona and the  companion star.   We first
consider   the corona, whose    output is  presumably   the result  of
upscattering  soft  disk  photons in  a  relatively hot  medium  (e.g.
Sunyaev and Titarchuk  1980).  For now,  the parameters  of this model
will be considered constant over time (excluding eclipses), except for
the normalization parameter which we will vary according to the linear
component of our light curve fit  from Sec.~\ref{sec:lc}.  This rather
false constraint  will   allow us to  separate  approximately  the two
spectral components  and is probably  valid across neighboring visits,
but not over a number of orbital cycles (Fig.~\ref{lcfit}).

The visit at  MJD~51372.15 is the  most likely to isolate the  coronal
component.   Eclipse   egress   is  approximately  complete   and  any
contribution from the reflection component is  assumed negligible.  We
refrain from  using  the pre-eclipse   visits at  MJD\,51\,373.4   and
MJD\,51\,375.2 in  case dipping is  occurring.   An absorbed power law
model of the form:
\begin{equation}
S(E) = \exp{\left[-N_{\mbox{\tiny H}}\sigma(E)\right]} A E^{-\alpha}
\end{equation}
where  $A$ is   the   normalization,  $\sigma(E)$  the  photo-electric
cross-section     (Balucinska-Church    and  McCammon      1992)   and
$N_{\mbox{\tiny H}}$ the  neutral hydrogen column density, provides an
inadequate fit of \chired\  =  2.06 for 54    dof. The addition  of  a
Gaussian model:
\begin{equation}
G(E) = \frac{A_{\mbox{\tiny K}}}{\sqrt{2 \pi \sigma^2_{\mbox{\tiny K}}}}
\exp{\left[-\frac{1}{2}\left( \frac{E -E_{\mbox{\tiny K}}}
{\sigma_{\mbox{\tiny K}}}\right)^2\right]}
\end{equation}
representing an iron emission complex at 6.4-6.7  keV improves the fit
to \chired\ = 1.15 for 51 dof.  $A_{\mbox{\tiny K}}$ is the line flux,
$\sigma_{\mbox{\tiny K}}$ the  line width and $E_{\mbox{\tiny K}}$ the
line center.  Compared to previous  observations of \her\ in anomalous
low-states (e.g. Parmar \etal\  1999; Coburn \etal 2000), the powerlaw
component is  rather steep, $\alpha  = 1.8 \pm 0.1$,  and the  Fe line
nested at  high energy $6.9   \pm  0.2$ keV.   By introducing  further
complexity into   the model both  the  powerlaw slope  and line energy
agree with previous    results  but  the $\chi^2$-statistic     cannot
discriminate  between a number of models.    A definitive model is not
essential  since these fit  parameters will remain  fixed  in order to
extract approximately  the  reflection  spectrum a few   visits later.
Therefore we choose a model which is  simple but physically realistic,
and is  also compatible with fits made   earlier in the  anomalous low
state by  Coburn  \etal  (2000)  --  in   this case  a  powerlaw  with
high-energy    exponential   cutoff, a    Fe emission    line   and  a
partial-covering, cold absorber:
\begin{equation}
S(E) = P(E) \left[ A 
\exp{\left(-\frac{E}{E_{\mbox{\tiny cut}}}\right)} E^{-\alpha} 
+ G(E) \right]
\label{eqn:model}
\end{equation}
where
\begin{equation}
P(E) = (1 - f) + f \exp{\left[-N_{\mbox{\tiny H}}\sigma(E)\right]}
\end{equation}
$f$  is the partial covering  fraction and  $E_{\mbox{\tiny cut}}$ the
cutoff energy.  The excessive number  of free parameters mean that the
column  density, covering   fraction   and powerlaw slope  are  poorly
constrained.  Since the current anomalous  low-state is thought to  be
the result of obscuration   rather than intrinsic source behaviour  we
can assume that the coronal photons  have undergone elastic scattering
and   rationalize fixing the powerlaw   slope  to it's main-high state
value,  $\alpha = 0.9$   (Dal Fiume \etal\ 1998).    From the visit at
MJD\,51\,372.15,   $N_{\mbox{\tiny   H}}  = (1.6    \pm 0.8$) \dex{23}
cm$^{-2}$, $f = 0.7 \pm  0.1$, $E_{\mbox{\tiny cut}}  = 11 \pm 2$ keV,
$A = (2.6 \pm 0.9) \dex{-3}$  photons cm$^{-2}$ s$^{-1}$ keV$^{-1}$ at
1 keV, $E_{\mbox{\tiny K}} = 6.7 \pm 0.1$ keV and $\sigma_{\mbox{\tiny
K}}  = 0.7 \pm  0.2$ keV with a line  strength of $(2 \pm 1) \dex{-4}$
photons cm$^{-2}$ s$^{-1}$.   $\chired  = 0.85$  for  50  dof  and the
equivalent width of the Fe complex is 775 eV.

We apply    this   same model   to    a visit   at   \phiorb\  =   0.5
(MJD\,51\,372.90).  Keeping $\alpha$  and $E_{\mbox{\tiny cut}}$ fixed
at their previous values provides a poor fit,  $\chired = 2.74$ for 51
dof.  Relaxing both   these constraints yields  an  acceptable  fit of
$\chired = 0.74$ for  49 dof.  However  $N_{\mbox{\tiny  H}} =  0$ and
$\alpha$ is extremely  flat,  $0.13  \pm 0.12$.   Consequently,  in  a
time-variable capacity the coronal model  alone does not fit the  data
sensibly. Therefore a two-source model is more appropriate.

\subsection{Reflection spectrum}
\label{subsec:reflection}

We would like  to test whether reflection provides  a good  fit to the
modulated component.  Reflection would be the  result of scattering of
hard  radiation from the   accreting source  off the  relatively  cold
atmospheres  of the  companion star  or accretion  disk.  The spectrum
will suffer photo-electric absorption at energies  less than $10$ keV,
have a strong Fe $K\alpha$ line and will be down-scattered at energies
higher than $50$ keV.  Thus, the reflection spectrum  has a maximum in
the \xte\ PCA band at energies $15-20$ keV.

As explained  in  the previous section, we   assume the  coronal model
parameters are  time-invariant, except for  the normalization which is
scaled by the light curve fit of Eqn.~\ref{eq:lc}. This is in order to
reduce  the  number of free    parameters which is  excessive  for the
quality of spectra from individual visits.

The visit we consider  is  at \phiorb\  $\simeq$ 0.5 (MJD  51\,372.90)
which has a number of advantages  over other observations.  This phase
provides the best counting   statistics.  Furthermore we expect  to be
observing $\sim$~100 percent of the irradiated surface at this time so
we  can assume that  all reflected  flux  is recorded. Finally we  are
reasonably close  to  the  visit  where   the  coronal spectrum    was
characterized and so   our   assumptions  concerning  the   form   and
brightness of this   component are  hopefully  valid.  We  test  these
assumptions in Sec.~\ref{sec:model}.

We first  attempt adding a few simple  models to the coronal component
in an   attempt to produce  a  statistically-acceptable fit.  A second
Gaussian component is always  required.  Adopting $S(E)$ + powerlaw  +
Gaussian provides a $\chired = 1.17$ for 52  dof.  The photon index is
flat, $\alpha = 0.33 \pm 0.1$, and the fit yields a notable low-energy
excess.  A $S(E)$  + blackbody + Gaussian yields  $\chired = 0.72$ for
52  dof.  The blackbody temperature is  $6.2 \pm 0.7$~keV and the $6.4
\pm 0.1$~keV  line has an equivalent width  of  3.13 keV.  Despite the
good fit,  it is  not  obvious what  would  radiate thermally  at this
temperature within the   framework of  the anomalous low-state   model
(Vrtilek  \etal 2001), or produce  such a broad  line.  The low-energy
excess remains a feature.

More realistically, we  fit $S(E)$  +  Gaussian  + the elastic,   cold
Compton  reflection model {\sc  hrefl}.  Although the geometry of this
model is  more suited to disk reflection  than a Roche  surface, it is
the  most-appropriate model within the {\sc  xspec} package and yields
an  approximate fit. This provides confidence  in a more sophisticated
approach  undertaken in  Sec.~\ref{sec:model}.  

The   fraction, $f_{\mbox{\tiny esc}}$, of  direct  X-rays observed is
zero, since  the  point source is   obscured  by the  accretion  disk.
However much of the companion star atmosphere has a largely-unobscured
view of   the  X-ray  source.    The   reflection  model   assumes  an
isotropically-radiating point  source of a   given energy spectrum and
flux that irradiates a structure subtending  a solid angle $\omega$ on
the sky with respect to the  point source.  By adopting stellar masses
of   M$_{\mbox{\tiny ns}}$  = 1.4\msun\  and  M$_{\mbox{\tiny com}}$ =
2.2\msun, and companion  radial velocity of 100\kmsec\ (Reynolds \etal
1997),  we  find $\omega  =   0.15\pi$.  We approximate   a reflecting
structure   incident     to    X-rays  at    all    angles     between
$\theta_{\mbox{\tiny min}} = 0$ degrees and $\theta_{\mbox{\tiny max}}
=  90$  deg, weighted equally, and  inclined  on  the sky at  an angle
$\theta_{\mbox{\tiny obs}} = 0$ deg to the  observer at \phiorb\ = 0.5
and orbital inclination $i  \simeq 90$ degrees.  Solar abundances were
adopted. The incident  energy spectrum is modeled  by the same cut-off
powerlaw as before with $\alpha = 0.9$ and $E_{\mbox{\tiny cut}} = 11$
\kev.

The  photon flux  of the irradiating   powerlaw spectrum and  Gaussian
profile provide  just four free  parameters,  yielding a  $\chired$ of
0.89 for 53  dof.  Fit parameters  are listed in Tab.~\ref{fits}.  The
fit is extremely  good considering  there  is only one free  parameter
available  to define the  reflection continuum.  Encouragingly, photon
flux for the   unseen powerlaw  source at 1   keV  is $(9.0 \pm   0.9)
\dex{-2}$   photons cm$^{-2}$ s$^{-1}$ keV$^{-1}$.    The  flux from a
typical  main-high state of   \her\  is 0.1 photons cm$^{-2}$   s$^{-1}$
keV$^{-1}$  at 1\kev\ (dal Fiume  \etal  1998).  The  line is  strong,
narrow and   nested  at low Fe   K$\alpha$  energies, as  predicted by
reflection models (Basko   \etal  1974; Ross \&  Fabian   1993).  This
consistency, combined with  the  period and  phasing of  the modulated
component all  point to  the companion  star providing the  reflective
surface.

\begin{deluxetable}{ll}
\footnotesize
\tablecaption{best-fit parameters for the three spectra presented
in  Fig.~\ref{spectra}.   All  fits   are  also  subject  to  a  fixed
interstellar column density of   $5.1 \dex{19}$ cm$^{-2}$   (Dal Fiume
1998).     The     reflection    parameters    are   given         for
MJD\,51\,372.90. \label{fits}}
\tablewidth{0pt}
\startdata
\hline\hline
Eclipse parameters & \\
\hline
$\alpha$                                     & $1.2 \pm 0.4$ \\
$A$ (ph cm$^{-2}$ s$^{-1}$ keV$^{-1}$)       & $(2 \pm 1) \dex{-4}$ \\
$E_{\mbox{\tiny K}}$ (keV)                   & $6.7 \pm 0.3$ \\
$\sigma_{\mbox{\tiny K}}$ (keV)              & $0.3 \pm 0.2$ \\
$A_{\mbox{\tiny K}}$ (ph cm$^{-2}$ s$^{-1}$) & $(6 \pm 3) \dex{-5}$ \\
\hline\hline
Coronal parameters & \\
\hline 
$N_{\mbox{\tiny H}}$ ($\dex{22}$ cm$^{-2}$)  & $16 \pm 12$ \\
$f$                                          & $0.7 \pm 0.6$ \\ 
$\alpha$\tablenotemark{a}                    & 0.9 \\
$E_{\mbox{\tiny cut}}$ (keV)                 & $11 \pm 2$ \\
$A$ (ph cm$^{-2}$ s$^{-1}$ keV$^{-1}$)     & $(3 \pm 1) \dex{-3}$ \\
$E_{\mbox{\tiny K}}$ (keV)                   & $6.7 \pm 0.2$ \\
$\sigma_{\mbox{\tiny K}}$ (keV)              & $0.7 \pm 0.4$ \\
$A_{\mbox{\tiny K}}$ (ph cm$^{-2}$ s$^{-1}$) & $(2 \pm 1) \dex{-4}$ \\
\hline\hline
Reflection parameters & \\
\hline 
$\alpha$\tablenotemark{a,b}                               & 0.9 \\
$E_{\tiny cut}$ (keV)\tablenotemark{a,b}                  & 11 \\
$A$ (ph cm$^{-2}$ s$^{-1}$ keV$^{-1}$)\tablenotemark{b} & $(9.0 \pm 0.9) \dex{-2}$ \\ 
$\omega$\tablenotemark{a}                                 & 0.15$\pi$ \\
$\cos{\theta_{\mbox{\tiny obs}}}$\tablenotemark{a}        & 1 \\
$\cos{\theta_{\mbox{\tiny min}}}$\tablenotemark{a}        & 1 \\
$\cos{\theta_{\mbox{\tiny max}}}$\tablenotemark{a}        & 0 \\
Fe abundance (solar)\tablenotemark{a}                     & 1 \\
$f_{\mbox{\tiny esc}}$\tablenotemark{a}                   & 0 \\
$E_{\mbox{\tiny K}}$ (keV)                                & $6.5 \pm 0.1$ \\
$\sigma_{\mbox{\tiny K}}$ (keV)                           & $0.1 \pm 0.1$ \\
$A_{\mbox{\tiny K}}$ (ph cm$^{-2}$ s$^{-1}$)              & $(1.5 \pm 0.2) \dex{-4}$ \\
\enddata
\tablenotetext{a}{Frozen parameter.}
\tablenotetext{b}{Model component is directly-invisible to the observer.}
\end{deluxetable}

\begin{figure*}
\begin{picture}(100,0)(10,20) 
\put(0,0){\includegraphics{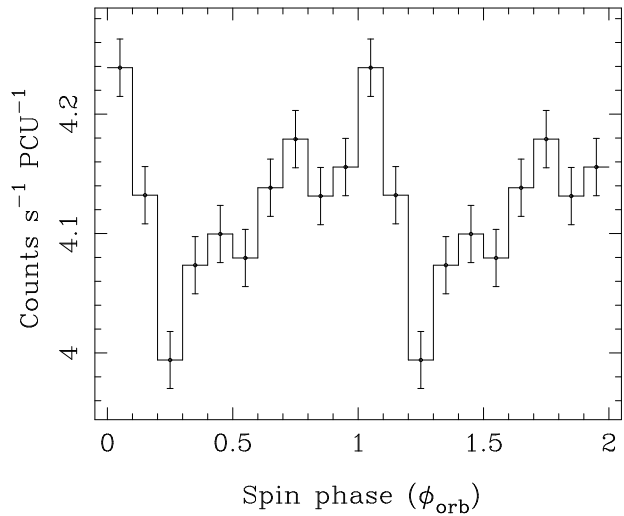}}
\noindent
\end{picture}
\vspace{50mm}
\figcaption[fig5.ps]{ Time-averaged spin pulse from \her. The pulse
cycle is repeated for clarity. \label{pulses}}
\end{figure*}

\begin{figure*}                          
\begin{picture}(100,0)(10,20)
\put(0,0){\includegraphics{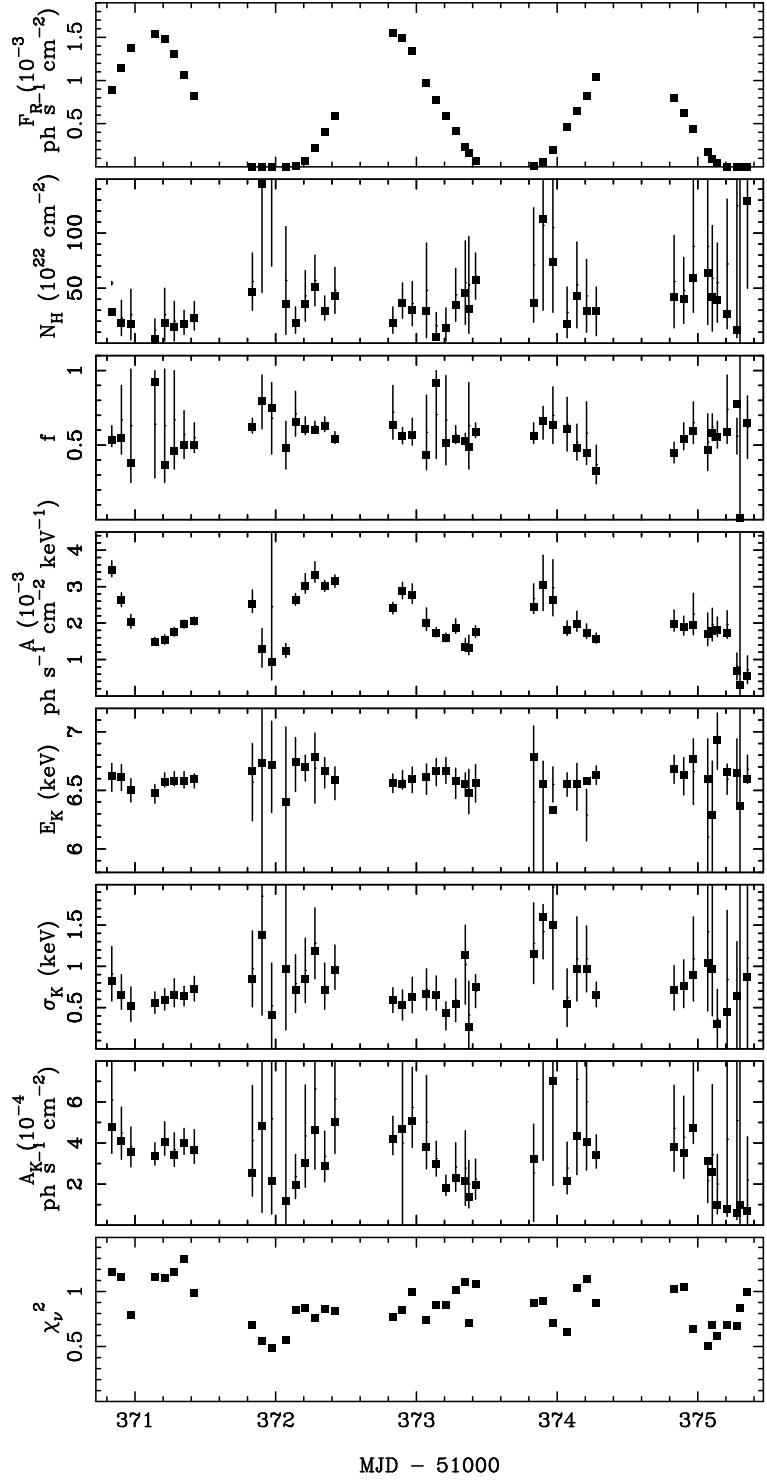}}   
\noindent  
\end{picture}   
\vspace{192mm}
\figcaption[fig6.ps]{The  top  panel presents the  predicted 3--30 \kev\ 
reflected flux from  the  companion star as  a function  of  MJD.  The
remaining  panels illustrate   the   fit  parameters  of  the  coronal
continuum model  and  a  single  Gaussian  fitting  both  coronal  and
reflected Fe    components after  the   reflection continuum  has been
subtracted.  The bottom panel provides  the  reduced $\chi^2$ for each
fit. \label{stelfit}}
\end{figure*}

\subsection{Eclipse spectrum}

The visit   during mid-eclipse (MJD\,51\,371.98)   is fit well  with a
single powerlaw,  $\chired  = 0.93$ for   26 dof over  channels  8--35
(3--15\kev).   Although the  inclusion of  a  Fe line is not  formally
required, there is  an excess longward of  the Fe K edge.  Including a
Gaussian model suggests a   line of equivalent  width 2.6   keV, where
$\chired = 0.39$ for 23 dof.  Here $\alpha = 1.2 \pm 0.3$, $A = (2 \pm
1)  \dex{-4}$   photons  keV$^{-1}$ cm$^{-2}$     s$^{-1}$ at  1  keV,
$E_{\mbox{\tiny K}} = 6.7 \pm 0.3$ keV  and $\sigma_{\mbox{\tiny K}} =
0.3 \pm 0.2$ keV with a line intensity of $(6 \pm 3) \dex{-5}$ photons
cm$^{-2}$ s$^{-1}$.   Adopting a cut-off powerlaw  model of  $\alpha =
0.9$   and   $E_{\mbox{\tiny cut}} = 11$\kev\    and  fitting just the
normalization    and line    parameters  yields   consistent  Gaussian
characteristics within the uncertainties.

\section{Pulse timing}

In this section we report the detection  of spin pulses, measure their
frequency  and present pulse profiles.   GoodXenon data were extracted
over all  energy channels and binned  to 0.02-s.  Times were corrected
to  the  compact  sources local standard  of  rest  using the  orbital
time-delay semi-amplitude  determined  by Deeter \etal (1991)  and the
spin  period of  the pulsar was  searched  for using the  Lomb-Scargle
algorithm.  A period  of $1.237746 \pm  0.000001$-s  was found with  a
false-alarm   probability of $10^{-5}$.  This  is  consistent with the
measurements of Parmar \etal (1999) and Coburn \etal (2000) during the
same  anomalous low-state and signifies that  the pulsar has spun down
during the low-state  episode, as it  has in the previous two recorded
anomalous low  epochs (Vrtilek   \etal   1994).  Coburn \etal   (2000)
discuss  the implications of these   spin-down episodes with regard to
the accretion torque in the system. The pulse profile is presented in
Fig.~\ref{pulses}.

It is unlikely  that the  observed  pulses are  within reflected light
from the  companion  star.  The   spin  frequency is  constant  in the
rest-frame of the   compact object so   should be  scattering off  the
accretion disk or accretion disk  corona. Pulse  fractions are of  the
order  a few  percent whereas  models   of reflected  pulses  off  the
companion predict fractions  an order  of magnitude less  (Middleditch
and  Nelson 1976).

\section{Modeling Roche lobe reflection}
\label{sec:model}

The current family of reflection  models available in {\sc xspec  v11}
are not  well-suited    to our problem.    These were   developed  for
accretion  disk reflection problems (see  e.g.  Magdziarz \& Zdziarski
1995) and limit our results because of incorrect geometry.  The models
available assume the reflecting  body is a  uniform slab, viewed at an
inclination angle and illuminated by a  point source above the surface
of the slab.   The size of the  slab and the  distance from  the point
source are characterized by  a filling factor  -- the fraction of  sky
occupied by the slab with respect to the source.  The model adopted in
Sec.~\ref{subsec:reflection}   takes   into  account     a  range   of
equally-weighted irradiance angles  to  approximate the curved   Roche
surface of the star, but  it is again a  simplification of the problem
(although a valuable consistency check for more sophisticated models).
The simplified model did indicate however that  the reflected flux was
identical (within   the 1-$\sigma$  measurement uncertainty)  to  that
expected  from a point source  with the distance, energy spectrum, and
intensity of \her\ at the peak of its main-high state.  We would like to
test this more rigorously by defining an accurate reflection surface.

In order to treat the problem more accurately  we develop a code which
divides the surface of a  Roche lobe-filling star into small, discrete
elements.  Each has an   individual inclination angle with  respect to
the Earth.  Each has an individual grazing incidence angle and filling
factor with respect to the  X-ray  source.  Our  task is to  determine
these   parameters over   a grid  using  Roche  geometry  and  sum the
reflection spectrum over  all observer-visible surface elements.  This
approach both treats  the  geometry correctly  and allows a  fit to be
made  at any orbital phase.   Again, we follow the algorithm described
in Appendix A.

The model  was  developed as an  external,  multiplicative {\sc xspec}
package and applied to  the spectrum sampled at  MJD 51372.90 where we
fit the composite model of Eqn.~\ref{eqn:model}, by replacing the {\sc
hrefl} component and adopting the  fixed coronal and pulsar parameters
listed in Tab.~\ref{fits}.  Our reflection parameters are stellar mass
ratio $q =   M_{\mbox{\tiny com}}/M_{\mbox{\tiny ns}}$,  which defines
the shape of   the companion star  and   the area of pulsar-sky   each
surface  element occupies,  the  orbital  inclination,  $i$,  and  the
orbital phase, \phiorb,  which together define  those elements  of the
stellar surface  visible    to   Earth.  We use     the  spectroscopic
measurements of Reynolds \etal (1997), $q  = 1.57$ and $i = 82^\circ$.
The   orbital ephemeris  of  Eqn.~\ref{eqn:orbit} indicates  \phiorb =
0.55.   In order to   calculate   the scattering  probability the   Fe
abundance is adopted as solar.

As before, four free parameters remain, yielding a reduced $\chi^2$ of
0.86 for  53  dof, where  $A  = (9.6 \pm  0.5)  \dex{-2}$  photons s$^{-1}$
cm$^{-2}$ keV$^{-1}$ at 1 \kev, $E_K =  6.5 \pm 0.1$ \kev, $\sigma_K =
0.1 \pm 0.1$  \kev\  and $A_K =  (1.4  \pm 0.2) \dex{-4}$ photons  s$^{-1}$
cm$^{-2}$.   This is  both consistent with  the  {\sc hrefl} model and
confirms  that the intrinsic flux  from the pulsar  is reproduced with
reflection off a Roche surface at this orbital phase.

At \phiorb\ = 0.5  the Roche reflection  model most-resembles the disk
reflection models.   Ideally we would like to  apply the test above to
all orbital  phases   but Fig.~\ref{lcfit} indicates   that if stellar
reflection drives the orbital  light curve, then the  coronal spectrum
is variable and our  model  assumptions break down.   The  alternative
approach is  to freeze $A$,  calculate the reflection spectrum at each
visit and then fit the coronal model to the residuals.

Since it  is not possible  to  separate the  coronal and reflected  Fe
lines, these are   fit with just   a single Gaussian.   Neglecting the
short-term behaviour of the absorbing column and coronal line flux, we
expect Gaussian  strength, $A_K$, and  width, $\sigma_K$, to vary with
the  reflection continuum as a  result of the narrow fluorescence line
from  the  companion star  varying over  the  orbital cycle. The 3--30
\kev\ reflection flux (top panel) and the  best fit parameters for all
visits are presented in Fig.~\ref{stelfit}.  Both the coronal $\alpha$
and $E_{\mbox{\tiny cut}}$ were frozen during the fits.

Coronal parameters  are expected to vary on  the 35-d, rather than the
1.7-d  cycle   and  we  see   no    detectable orbital coherency    in
$N_{\mbox{\tiny  H}}$, $f$ or $A$.    The  decay of  the coronal  flux
appears related to    a long-term increase    in $N_{\mbox{\tiny H}}$,
rather than a decrease in $A$, which is  consistent with the 35-d disk
precession model  for \her\ (c.f. Scott  \& Leahy 1999). As predicted,
$A_K$ is modulated  on the orbital period, with  a maximum at \phiorb\
$\simeq$ 0.5,  although   similar  coherency is  not    discernible in
$\sigma_K$.

\section{Discussion}

Despite the reflection effect in \her\ being  observed in the EUV band
(Leahy  \&  Marshall  1999;  Leahy  \etal   2000),  this is the  first
convincing detection of the phenomenon at X-ray energies.  Basko \etal
(1974) demonstrated that X-ray reflected  flux from the companion star
in  \her\  should be  at  the level  a  few  percent of  the intrinsic
radiation.  This should generally be washed out by shot noise from the
central source during  normal main-high states,   but as in  the current
observations,  should also be observable during  normal  low states of
the 35-d cycle.  Sheffer \etal (1992) claimed a tentative detection of
reflection during a normal  off-state during the {\it ASTRON} mission,
but  a systematic  search for  this  effect during normal or anomalous
low-states has never been undertaken;  the HEASARC archive is not rich
with  data from  these  epochs  and those  that   exist coincide  with
eclipses of the pulsar (e.g. Choi \etal 1997). Exceptions are the {\it
EXOSAT}  pointings  by  Parmar  \etal    (1985),  the {\it   BeppoSAX}
observations of Parmar \etal (1999) and the \xte\ visits by Coburn
\etal  (2000).  None  of these    analyses  consider reflection as   a
possible  interpretation of data,  but   the count rates presented  by
Parmar \etal (1985; 1999) arguably both display 1.7-d modulation.  The
baseline of the Coburn    \etal  (2000) visits is    20~ksec  covering
\phiorb\  =  0.64--0.77.  The general  trend  in count  rate over this
period is  one  of  decline, as predicted   by  the  reflection  model
(Fig.~\ref{lcfit}),  although the count rate  and decline are a factor
two larger than the current data.

Potentially, \her\  contains a rich  lode of reflection  physics.  The
reflection effect  is  common to a   large number of   different X-ray
sources such  as AGN and  stellar mass black  hole (Lightman \& White;
White \etal 1988).  But generally  there is uncertainty concerning the
physical location and size of hard radiation sources in these objects,
and the size and shape  of the reflecting  atmosphere (which in  these
cases is presumably an accretion disk).  Therefore Her X-1 is an ideal
X-ray source for studying  the properties of reflection and associated
processes (such as winds and coronae etc) because we have a relatively
clear   picture of  companion  star geometry   and its illumination by
well-measured hard radiation from a central object.

Despite this there are some limitations.  Firstly it is not clear how,
or  whether,  the X-ray  emission from the   pulsar is  beamed, or how
column density changes with disk inclination angle.  Although the high
inclination  of the binary means  the observer has  a  similar line of
sight to  the source as the  equatorial regions of the companion star,
surface elements closer to the stellar poles  may perceive the central
source differently.  In this  paper we  have assumed the  source is an
isotropic emitter. Also we assume that the X-ray source is point-like.
This is not true, since we observe scattered  light from a disk corona
and expect  further scattering  and  reflection from the  disk itself;
Leahy   (1995) showed that soft  photons  come from an extended region
using  eclipse timings.  Furthermore we made  no  attempt to model the
accretion disk shadow, or its eclipse, over  the equatorial regions of
the companion star since the disk shape is so poorly constrained.  The
reflection process itself is also  treated simply.  More sophisticated
models by (e.g.   Ross \& Fabian  1993;  Nayakshin 2000) consider  the
physical  response of an atmosphere  to irradiation  and the resultant
spectral changes.

There is scope for much more observation.  The reflection model can be
tested more sensitively with  improved    models and greater    energy
resolution using  grating spectrometers   sensitive to  soft  energies
below 1 \kev\ which should  be populated by resonance absorption edges
(Basko \etal  1974), and over the  Fe\,$\alpha$ complex to resolve and
separate line components between corona  and companion star.  In order
to  constrain  the  location  of reflection,  timing  experiments with
improved sampling over  the 35-d cycle  would determine  whether light
curve   modulation is driven  on the  orbital (1.7-d)  or beat (1.6-d)
period.   Fast timing visits during eclipses  of  the companion by the
accretion disk at \phiorb\ = 0.5  can help constrain  the shape of the
accretion disk using the eclipse ingress and egress profiles.

\section{Conclusion}

Over  2.7 orbital cycles during  the  anomalous low-state of 1999  the
X-ray  pulsar  \her\ displayed an X-ray   light curve very  similar to
optical    and UV counterparts,   suggesting  that  the companion star
contributes to X-ray emission.  A cold Compton reflection spectrum was
successfully fit to the variable component of the energy spectrum.  By
employing binary geometry and  assuming companion star reflection, the
intensity  of  the hidden pulsar  was  found  to  be  identical to the
main-high state of the source when the  pulsar is visible. This result
indirectly    implies  the anomalous  low-state   model  is correct --
intrinsic accretion luminosity remains constant, but is occulted along
the observers line-of-sight by the accretion disk.

\acknowledgments
This work  was funded by  NASA grant  NRA-99-01-ADP-108, NAG5-6711 and
NAG5-7333. KO, KH and HQ acknowledge support from the Particle Physics
and Astronomy Research  Council. This research   has made use  of data
obtained  from the High Energy   Astrophysics Science Archive Research
Center (HEASARC), provided by NASA's Goddard Space Flight Center.

\appendix

\begin{figure*}
\begin{picture}(100,0)(10,20) 
\put(0,0){\includegraphics{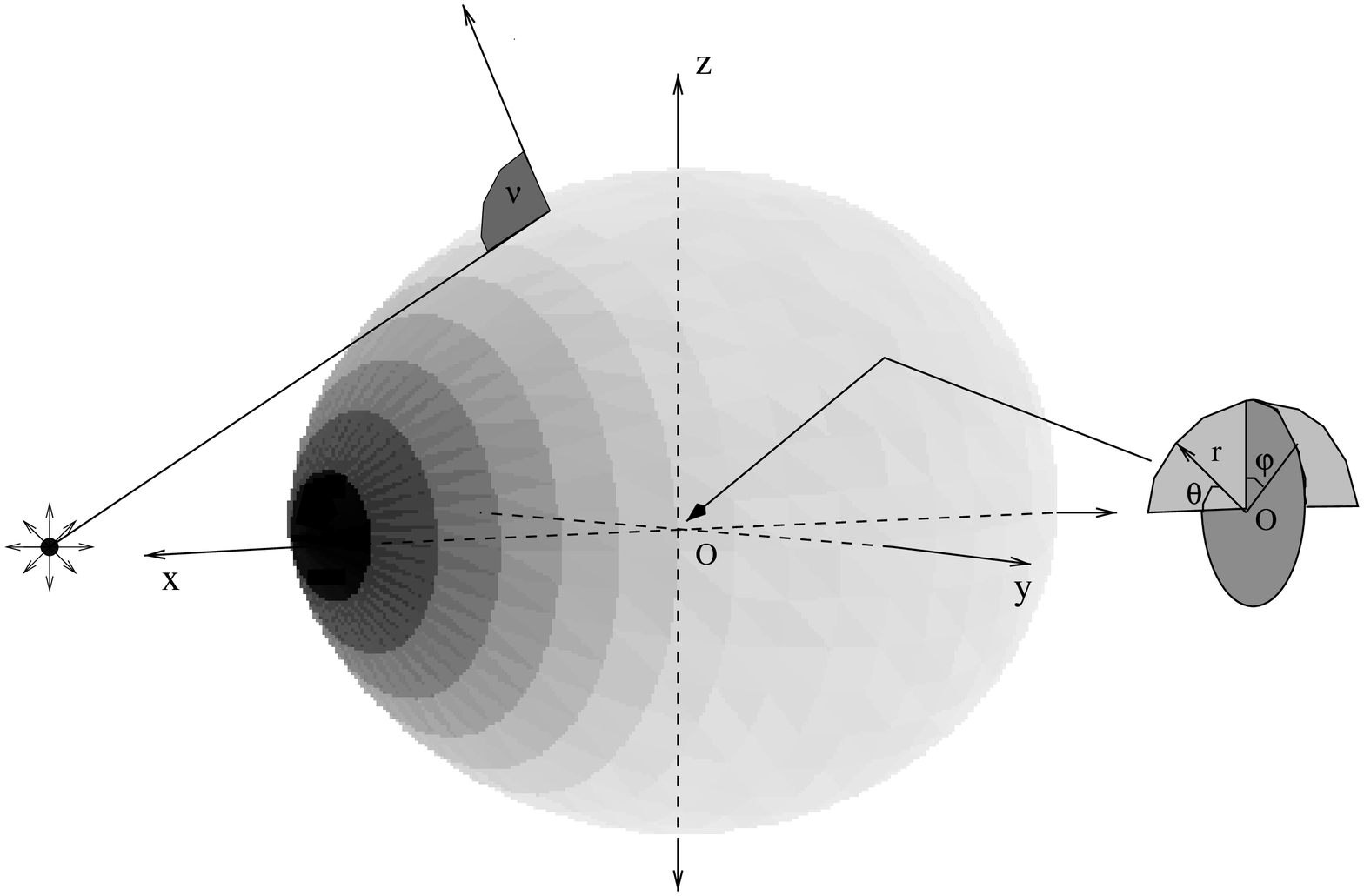}}
\noindent
\end{picture}
\vspace{50mm}
\figcaption[fig7.ps]{Schematic of the geometry used to determine the 
reflected  spectrum,    both   cartesian and  polar    coordinates are
displayed.  $\nu$ is the incidence angle of reflection relative to the
neutron  star.  The incidence  angle between surface  elements and the
observers  line of  sight,  $\mu$,  is the  projection   angle of each
surface element on the page.\label{schematic}}
\end{figure*}

\section{Classical Compton reflection off a Roche lobe-filling surface}

In order   to estimate the  stellar-reflected flux  one  must know the
fraction of X-rays intercepted by the atmosphere  of the companion and
also the reflection   properties of  that  atmosphere.  The  following
calculations construct a   stellar  surface  confined by   the   Roche
equipotential surface   of  a binary  of  stellar   mass  ratio  $q  =
M_{\mbox{\tiny   com}}/M_{\mbox{\tiny ns}}$    (Sec.~\ref{app:roche}).
They then determine the  monochromatic Compton reflected fraction  off
discrete surface elements and sum the reflected flux in those elements
visible to the observer at binary inclination $i$ and binary phase
\phiorb (Sec.~\ref{app:spectrum}).    

\subsection{Roche geometry}
\label{app:roche}

The critical Roche potential is defined by: 
\begin{equation}
\Psi_{\mbox{\tiny cr}} = \frac{q}{r_{\mbox{\tiny cr}}} +
\frac{1}{1 - r_{\mbox{\tiny cr}}} + 
\frac{(1 + q)r_{\mbox{\tiny cr}}^2}{2} 
\label{eqn:psicri}
\end{equation}  
$r_{\mbox{\tiny  cr}}$ is  normalized   to the stellar  separation and
determined   by  minimizing    $\left|  \epsilon_{\mbox{\tiny cr}}   /
(\mbox{d}\epsilon_{\mbox{\tiny  cr}}/\mbox{d}r_{\mbox{\tiny      cr}})
\right|$, where
\begin{equation}
\epsilon_{\mbox{\tiny cr}} =
(1 + q)r_{\mbox{\tiny cr}} - 1 - 
\frac{q}{r_{\mbox{\tiny cr}}^2} + 
\frac{1}{(1 - r_{\mbox{\tiny cr}})^2}
\end{equation} 

Using a spherical  polar coordinate system ($r$,$\theta$,$\phi$), with
an origin  at the center of mass  of  the companion star,  the stellar
surface is  divided over $\theta$  and  $\phi$ into segments of  equal
solid angle (see   Fig.~\ref{schematic}). For any given  $\theta$  and
$\phi$,   $r$   is  determined    by    minimizing $\left|\epsilon   /
(\mbox{d}\epsilon/\mbox{d}r)\right|$, where
\begin{equation}
\epsilon = 
\frac{q}{r} + 
(1 + r^2 -2r\cos{\theta})^{1/2} +
\frac{(1 + q)r^2 \cos{^2\theta}}{2} +
\sin{^2\theta} \cos{^2\phi} +
\frac{1}{2(1 + q)} -
r\cos{\theta} -
\Psi_{\mbox{\tiny cr}}
\label{eqn:rho}
\end{equation}

Following the approach previously adopted by Rutten and Dhillon (1994)
and Still \etal (1997), each surface segment is divided into two, such
that the critical Roche surface  is completely tiled by $N$ triangles.
This provides a better  approximation to a  curved surface  than other
polygons.  The coordinates of each  triangle vertex, {${\bf V}_{jk}  =
(r_{jk},\theta_{jk},\phi_{jk})$}, where $j$  = 1 to  3 and $k$ = 1  to
$N$, are  calculated  using  Eqns.~\ref{eqn:psicri}--\ref{eqn:rho} and
converted  to a  Cartesian coordinate system  ($x_{jk},y_{jk},z_{jk}$)
with an origin at the center of mass using:
\begin{equation}
x_{jk} = r_{jk}  \cos{\theta_{jk}} - \frac{1}{1  + q}
\end{equation}
\begin{equation}
y_{jk} =  r_{jk} \sin{\theta_{jk}} \cos{\phi_{jk}}
\end{equation}
\begin{equation}
z_{jk}  = r_{jk} \sin{\theta_{jk}} \sin{\phi_{jk}}
\end{equation}
The    coordinate  center   of   each  triangle  is   ${\bf   T}_k$  =
$\sum_{j=1}^3(x_{jk}/3,y_{jk}/3,z_{jk}/3)$.  The unit vector normal of
each triangular surface is
\begin{equation}
\boldmath{n}_k = \frac{(\boldmath{V}_{2k} - \boldmath{V}_{1k}) \times
(\boldmath{V}_{3k}  - \boldmath{V}_{1k})}{\left|(\boldmath{V}_{2k} - 
\boldmath{V}_{1k}) \times (\boldmath{V}_{3k}  - 
\boldmath{V}_{1k})\right|} 
\end{equation}

where the surface area of each element $A_k = \left|{\bf n}_k\right| /
2$.    The cosine of a surface   elements normal  vector  and the unit
direction   vector to  Earth, ${\bf  e}$,   is the incidence angle  of
reflected photons:
\begin{equation}
\mu_k = \boldmath{n}_k . \boldmath{e}
\end{equation}
${\bf e}$ is defined using 
the orbital inclination  of the binary,  $i$,  and the orbital  phase,
\phiorb.
\begin{equation}
e_x = \sin{i} \cos{(\pi (1 - 2 \phiorb))}
\end{equation}
\begin{equation}
e_y = \sin{i} \sin{(\pi (1 - 2 \phiorb))}
\end{equation}
\begin{equation}
e_z = \cos{i}
\end{equation}
Similarly   the cosine of each elements   normal  vector with the unit
vector direction  to the X-ray  source yields  the  incidence angle of
photons arriving at the stellar surface:
\begin{equation} 
\nu_k = \boldmath{n_k} . \frac{\boldmath{d_k}}{\left|d_k\right|}  
\end{equation}   
where 
\begin{equation}
d_{x,k} = - \boldmath{T}_{x,k} +  \frac{q}{1 + q}
\end{equation}
\begin{equation}
d_{y,k} = - \boldmath{T}_{y,k}
\end{equation}
\begin{equation} 
d_{z,k} = - \boldmath{T}_{z,k}
\end{equation}
and $\left|d_k\right|$  is  the distance  between the source  and each
triangle. We assume the X-ray  source is point-like and is  coincident
with the center of mass of the pulsar.  Finally the solid angle of sky
subtended by each triangle with respect to the X-ray source is
\begin{equation}
\omega_k = \nu_k 
\left[
1 - \cos{\left(\tan{^{-1}
\left(\frac{\sqrt{A_k / \pi}}{\left|{\boldmath d}_k\right|}\right)}
\right)}\right]
\end{equation}

\subsection{Reflected flux}
\label{app:spectrum}

In Chakrabarti and Titarchuk (1995) the reflection  problem in a plane
geometry was solved  exactly  using  the Fokker-Plank   treatment  for
multiple    scattered     photons.     Photoelectric absorption    and
downscattering effects were taken into account correctly. They suggest
that  in  the  very  non-relativistic  energy  range,  $<$~30  keV the
downscattering effects are negligible.  Photons scatter  off electrons
almost coherently  and   consequently  the results  of  the  classical
reflection problem   are  applicable (Chandrasekhar   1960).   If  the
intrinsic  flux spectrum from a  point  source radiating isotropically
from the location of the pulsar is  $F_E$, the reflected flux observed
off all surface elements is
\begin{equation} 
f_E  = \frac{\lambda_E F_E}{2} 
\sum_{k=1}^N{\frac{\omega_k \mu_k}{\mu_k + \nu_k}}
H(\lambda_E,\mu_k) H(\lambda_E,\nu_k)~~~~~~~(\mu_k > 0, \nu_k > 0)
\end{equation} 
Unobservable  elements with $\mu_k <  0$ and elements  not incident to
photons from the pulsar, $\nu_k < 0$,  should be removed from the sum.
$\lambda_E$ is the  photon scattering probability approximated  by the
piecewise    polynomial    fit  of    Morrison  \&    McCammon (1983).
$H(\lambda_E,\beta)$ is the classical H-function (Chandrasekhar 1960),
approximated with an   uncertainty  of $<$  1\%  using  the analytical
expression of Basko (1978):
\begin{equation}  
H(\lambda_E,\beta) =
\frac{1 + \sqrt{3} \beta}{1 + \sqrt{3(1 - \lambda_E)}\beta} 
\left[ 
1 - \frac{\lambda_E \beta}{4} 
(1 + \lambda_E^2) (\ln{\beta} + 1.33 - 1.458 \beta^{0.62}) 
\right] 
\end{equation}

\end{document}